\documentclass[twocolumn]{aastex7}
\usepackage{amsmath,amstext}
\usepackage[T1]{fontenc}
\usepackage{apjfonts}
\usepackage[figure,figure*]{hypcap}
\usepackage{multirow}
\usepackage{subfiles}
\usepackage{graphicx}
\usepackage{subcaption}
\usepackage{xspace}
\DeclareGraphicsExtensions{.pdf,.png}

\newcommand{\Aeos}{A{\sc eos}\xspace}

\defcitealias{Emerick2019}{E19}
\defcitealias{AeosMethods}{B24}

\begin{document}

\title{\textsc{Aeos} is Mixing it Up: The (In)homogeneity of Metal Mixing Following Population III Star Formation}

\correspondingauthor{Jennifer Mead}
\email{jennifer.mead@columbia.edu}


\author[0009-0006-4744-2350]{Jennifer Mead}
\affiliation{Department of Astronomy, Columbia University, New York, NY 10027, USA}
\email{jennifer.mead@columbia.edu}

\author[0000-0002-8810-858X]{Kaley Brauer}
\affiliation{Center for Astrophysics | Harvard \& Smithsonian, Cambridge, MA 02138, USA}
\email{kaley.brauer@cfa.harvard.edu}

\author[0000-0003-2630-9228]{Greg L. Bryan}
\affiliation{Department of Astronomy, Columbia University, New York, NY 10027, USA}
\affiliation{Center for Computational Astrophysics, Flatiron Institute, 162 5th Ave, New York, NY 10010, USA}
\email{greg.bryan@columbia.edu}

\author[0000-0003-0064-4060]{Mordecai-Mark Mac Low}
\affiliation{Department of Astrophysics, American Museum of Natural History, New York, NY 10024, USA}
\affiliation{Department of Astronomy, Columbia University, New York, NY 10027, USA}
\email{mordecai@amnh.org}

\author[0000-0002-4863-8842]{Alexander P. Ji}
\affiliation{Department of Astronomy \& Astrophysics, University of Chicago, 5640 S Ellis Ave, Chicago, IL 60637, USA}
\affiliation{Kavli Institute for Cosmological Physics, University of Chicago, Chicago, IL 60637, USA}
\email{alexji@uchicago.edu}

\author[0000-0003-1173-8847]{John H. Wise}
\affiliation{Center for Relativistic Astrophysics, School of Physics, Georgia Institute of Technology, Atlanta, GA 30332, USA}
\email{jwise@physics.gatech.edu}

\author[0000-0003-3479-4606]{Eric P. Andersson}
\affiliation{Department of Astrophysics, American Museum of Natural History, New York, NY 10024, USA}
\email{eandersson@amnh.org}

\author[0000-0002-2139-7145]{Anna Frebel}
\affiliation{Department of Physics and Kavli Institute for Astrophysics and Space Research, Massachusetts Institute of Technology, Cambridge, MA 02139, USA}
\email{afrebel@mit.edu}

\author[0000-0003-2807-328X]{Andrew Emerick}
\affiliation{Carnegie Observatories, Pasadena, CA 91101, USA}
\email{aemerick11@gmail.com}

\author[0000-0002-9986-8816]{Benoit C{\^o}t{\'e}}
\affiliation{Department of Physics and Astronomy, University of Victoria, Victoria, BC, V8P5C2, Canada}
\email{cotebenoit8@gmail.com}


\begin{abstract}
Stellar surface abundances are records of the state of the gas from which stars formed, and thus trace how individual elements have mixed into the surrounding medium following their ejection from stars.  In this work, we test the common assumption of instantaneous and homogeneous metal mixing during the formation of the first Population II stars by characterizing the chemical homogeneity of the gas in simulated star-forming environments enriched by Population III stellar feedback. Testing the homogeneity of metal mixing in this time period is necessary for understanding the spread of abundances in the most metal-poor stars, and the (in)homogeneity of individual sites of star formation.  Using \Aeos, a suite of star-by-star cosmological simulations, we quantify how gas abundances change over space and time relative to Population II stellar abundances using Mahalanobis distances, a measure of covariance-normalized dissimilarity. We find that the homogeneous mixing assumption holds only within $\sim100$ pc of a star-forming region and $\sim 7$ Myr following the star formation event. Beyond this regime, deviations between stellar and gas abundances increase until they become indistinguishable from assuming a homogeneous mix of metals averaged over the initial mass function. This highlights the limited applicability of assuming instantaneous and homogeneous mixing in realistic halo environments at high redshift.  We identify critical mixing scales that are necessary to explore chemical evolution in the early Universe. These scales can be applied to determine the precision needed for accurate chemical tagging of observed data and to explore parameter space with analytical models.
\end{abstract}

\section{Introduction}
Chemical evolution is a consequence of stellar evolution and nucleosynthesis, encoding properties such as the initial mass function and star formation history of galaxies, and directly shaping stellar abundances.  Core-collapse supernovae (CCSNe), Type Ia supernovae (SNIa), stellar winds, and other feedback processes return newly synthesized elements to the interstellar medium (ISM), where they can be incorporated into future generations of stars.  Tracking the build-up of elements over space and time with analytic and numerical galactic chemical evolution models provides the link between theoretical nucleosynthetic yields and global galactic properties. These include the metallicity distribution function, abundance gradients, and trends such as ${\rm [\alpha/Fe]}$ vs.\ ${\rm [Fe/H]}$.\footnote{[X/Fe] is the log of the fractional number density of element X to the number density of Fe relative to the solar abundances.  $\rm [X/Fe] = log(n_{X,*}/n_{Fe,*}) - log(n_{X,\odot}/n_{Fe,\odot})$}

A key ingredient in these models is how elements get mixed across the ISM. However, the details of chemical mixing into the ISM have long been poorly treated or essentially ignored in analytical models. The assumption of instantaneous and homogeneous mixing across galaxies dominates due to its ability to largely reproduce the aforementioned large-scale relations \citep[e.g.,][]{Tinsley1980,Matteucci1986,Matteucci1989,Chiappini1997,Spitoni2010}.  Some works have made more detailed models of chemical evolution by breaking down the galaxy spatially into smaller one zone models \citep[e.g.,][]{Boissier1999,Spitoni2011,Hegde2025}, adding a multiphase ISM \citep[e.g.][]{Samland1997,Ploeckinger2014}, or parameterizing the volumes or masses over which metals are homogeneously mixed over time or the variation in metallicities of star forming gas \citep[][]{Karlsson2008,deBennassuti2014,deBennassuti2017,Hartwig2018,Salvadori2019,Tarumi2020b}.  

Until the last decade or so, without high-precision abundance observations, instantaneous and homogeneous mixing has been sufficient to explain large-scale galactic relations at metallicities [Fe/H] $\geq -2$.  However, evidence for inhomogeneous mixing has long been observed in the most metal-poor stars via the neutron-capture elements \citep[e.g.][]{Ryan1996,McWilliam1997,Francois2003}. More recently, theoretical models in tandem with large-scale surveys of high-precision abundances, have also suggested that metal mixing is inhomogeneous at higher metallicities, or at least that the homogeneity is scale dependent \citep[e.g.,][]{Greif2010,PanScannapiecoScalo2013,FengKrumholz2014,Ritter2015,Hill2019,Ji2020,Mead2024}.

Inhomogeneous metal mixing over galactic scales, on the other hand, has also been invoked to explain the lack of observations of pair instability SNe signatures in metal-poor stars \citep[][]{Magg2022}, the formation of carbon-enhanced metal-poor stars \citep[][]{Sarmento2017,HartwigYoshida2019,Magg2020}, and extended Population III (Pop III) star formation \citep[][]{Tornatore2007,Sarmento2018,LiuBromm2020,Hegde2025}.  \citet{Ji2015} showed it is possible that Pop III chemical signatures may survive in the most metal-poor stars in the halo due in large part to inhomogeneous metal mixing, although those signatures would quickly become diluted in the gas with subsequent Population II (Pop II) chemical yields.  Conversely, on small scales, metal mixing by turbulence is argued to be efficient enough to explain the homogeneity observed in open clusters \citep[][]{FengKrumholz2014}.

Observationally, inhomogeneities can be identified by scatter in stellar \citep[e.g.,][]{Venn2012,Hill2019,Ji2020,Mead2024} and gas \citep[e.g.,][]{Kreckel2020} abundances.  Using the PHANGS-MUSE survey of eight nearby galaxies, \citet{Kreckel2020} identified greater homogeneity (lower scatter; 0.02--0.03 dex) in $\Delta$O/H gas abundances on scales $<600$ pc than globally (0.03--0.05 dex).  In simulations, these spreads have been attributed to the inefficiency of metal mixing \citep[][]{Ferrara2000}.

In theory and simulations, attempts to identify the spatial and temporal scales of mixing have yielded varied results.  These studies have focused on several aspects that drive the homogenization of metals including metal injection timing, geometry, sources, turbulence, and shear \citep[e.g.,][]{deAvillez2002,Bland-Hawthorn2015,Petit2015,Hirai2017-1,Escala2018,Emerick2020a,Kolborg2022,Hirai2024,Sharda2024,Zhang2025}.  Spatial mixing scales range from sub-kpc \citep{deAvillez2002} to kpc scales \citep{KrumholzTing2018,Zhang2025} for Milky Way-like galaxies, and on the order of tens to hundreds of parsecs for dwarf galaxies and Milky Way progenitors at high redshift \citep{KrumholzTing2018,Kolborg2022}.  \citet{Corlies2018} found that high-redshift dwarf galaxy progenitors demonstrate gas metallicities within 50\% of the virial radius that agree with a closed box model, motivated by the assumption that metals are retained and that there is no inflow or outflow of gas and metals. However, this assumption breaks down at the outskirts of galaxies due to inflows and outflows, and the metals are unlikely tp be described as well-mixed at any given radius or at any given time.

Mixing timescales can range anywhere from 100--350 Myr for Milky Way-like galaxies \citep{deAvillez2002,Yang2012,KrumholzTing2018,Zhang2025} to just tens of Myr for dwarf galaxies \citep{Hirai2017-1,KrumholzTing2018}, although \citet{Emerick2020a} found timescales of 100 Myr–-1 Gyr for a dwarf galaxy, with a dependence on the source of enrichment.  \citet{Petit2015} demonstrated that the decay time for inhomogeneities due to turbulence and shear largely depends on the symmetry of the mode, with axisymmetric modes dissipating inhomogeneities over longer timescales, thus explaining the existence of radial metallicity gradients and lack of azimuthal gradients in most galaxies.

The exact mechanisms driving the spatial and temporal scales of mixing also remain controversial.  Early work \citep{deAvillez2002} found that a primary driver of mixing time is the rate of supernovae (SNe) over some area, with higher SN rates leading to shorter mixing timescales, but that, for a given rate, the timescale for mixing was independent of the length scale of the inhomogeneity.  Similarly, \citet{Kolborg2022} indirectly explored the impact of the SN rate via the star formation rate. They found that lower star formation rates were associated with more homogenized media at the same mean metallicity due to turbulence driving homogenization, suggesting lower mixing timescales as the media were not subject to constant injection of new metals.  In contrast, \citet{Zhang2025} explored the injection of several isotopes from a variety of sources. They argued that the correlation length between element abundances, or the scale over which abundances remain correlated and thus not well-mixed, is a result of the timing of injection and stellar drift.  Thus, metals injected on short timescales have shorter correlation lengths than those injected on longer timescales where conatal stars have drifted apart.  \citet{Emerick2020a} found that individual enrichment sources that inject their metals with higher energies (e.g., SNe) display greater homogeneity both spatially and temporally.  Metal mixing in the ISM has also been studied across gas phases, though here, it is widely agreed upon that hot gas mixes more efficiently than cold gas \citep[e.g.,][]{KobulnickySkillman1997,deAvillez2002,Emerick2018b,Emerick2019,Emerick2020a}.

The spatial and temporal scales over which metals are mixed has implications for reconstructing star formation histories.  For example, the premise of chemical tagging \citep{Freeman2002}---the idea that stellar birth siblings can be identified through stellar abundances---rests heavily on the assumption that the gas from which birth siblings form is well-mixed, and that observational abundances have high enough precision to access the chemical space dimensionality required to trace back spatially disjoint stars to uniquely differentiated birth sites. In this case, metal mixing must occur on scales large enough such that stellar siblings have similar enough abundances to be linked, and simultaneously small enough scales to differentiate from other sites of star formation.

Many studies on spatial and temporal mixing scales seek to quantify at what point metals that are injected through feedback events are distributed uniformly across some volume.  In this work, we seek to answer a similar, but fundamentally different question: how valid is the assumption of instantaneous and homogeneous mixing \textit{at the time of star formation}?  The framing of the question in this way is particularly important for chemical tagging, as it matters not so much how long metals from feedback take to mix into the ISM, but rather over what spatial and temporal scales a star formation event can occur such that we expect stellar siblings to have similar enough abundances to be identified.

We test the instantaneous and homogeneous mixing assumption by quantifying the chemical (in)homogeneity of gas relative to stars in \Aeos, a suite of star-by-star cosmological simulations of the first stars and galaxies.  After Pop III stars form from metal-free primordial gas, we focus on the Pop II stars that form shortly after Pop III supernovae, because they form in a relatively clean chemical environment, enriched by only a few sources, and are therefore ideal probes of early mixing (and are more likely to survive long enough to be observed). The \Aeos simulation suite \citep{AeosMethods} is especially suited for studying metal mixing in the early universe for three reasons. First, it uses high spatial resolution, which allows detailed tracking of the spatial evolution of abundances. Second, it models individual stars, which enables individual stellar feedback and injection of metals at discrete locations from mass and metallicity dependent yields. Finally, it focuses on Pop III and early Pop II star formation, which provides a clean environment to track metal mixing. Using the \citet{Mahalanobis1936} distance as a multidimensional metric of abundance deviation, we measure how chemically distinct early Pop II stars are from their local gas reservoirs across a range of spatial scales and across time following formation.  By comparing stellar abundances to the average gas abundance within different spatial volumes and time frames, we assess how well the instantaneous and homogeneous mixing assumption captures the actual conditions of star-forming gas.

This paper is organized as follows. In Section \ref{sec:sim} we summarize the key features of the \Aeos simulations that are relevant to this study.  In Section \ref{sec:MD} we detail our metric for quantifying the (in)homogeneity of mixing in the ISM, and follow this in Section \ref{sec:results} with a study of instantaneous and homogeneous mixing over different spatial and temporal scales, as well as different phases of gas.  We conclude and discuss the implications of our results, particularly for chemical tagging initiatives, in Section \ref{sec:conc}.

\section{Simulation} \label{sec:sim}
The \Aeos simulations \citep{AeosMethods} are a suite of 1~cMpc$^3$ star-by-star simulations that model individual stellar feedback with high spatial and dark matter mass resolution (1 physical parsec and 1840 $M_\odot$, respectively). \Aeos is run using a modified version of the adaptive mesh refinement cosmological hydrodynamics code ENZO \citep{Enzo2014,Enzo2019,Emerick2019}, with cosmological parameters from the \citet{Planck2014} best fit.  With its spatial resolution, \Aeos resolves all minihalos down to a halo dark matter mass $M_{\rm dm} \sim 10^5 \, M_\odot$ (i.e. well below the expected minimum mass for Pop III star formation).  We restrict our description of \Aeos here to a summary of the relevant physics and setup; a complete description of the simulation can be found in \citet{AeosMethods}.

\subsection{Star Formation}
Stars are formed in cells containing at least 100 $M_\odot$ of gas that is converging, with $\nabla \cdot v < 0$, cold, with $T<500$~K, and dense, with $n>10^4\, {\rm cm^{-3}}$. Masses for each star are drawn stochastically from an adopted initial mass function (IMF) until the gas reservoir is depleted.  Each star formed acquires the elemental abundance of the gas within the cell in which it forms.

\subsubsection{Pop III Star Formation}
Pop III star formation has the additional constraint that gas also must have a molecular hydrogen fraction $f_{H_2} > 5\times10^{-4}$.  Pop III stars are distinguished from Pop II stars by having metallicity $Z<10^{-5}\,Z_\odot$.  For Pop III stars, we adopt an IMF similar to \citet{Salpeter1955} with a power-law slope of $\alpha=-1.3$ above a characteristic mass $M_{\rm char}$ and an exponential cutoff below this threshold.  In this work, we use the Aeos20 simulation \citep[described in][]{Brauer2025b, Mead2025a}, which uses $M_{\rm char} = 20 \, M_\odot$, $M_{\rm min} = 1\,M_\odot$, and $M_{\rm max} = 300 \, M_\odot$ for Pop III stars.  All Pop III star particles represent individual stars. Stellar lifetimes are drawn from \citet{Schaerer2002}.

\subsubsection{Pop II Star Formation}
Pop II stars are formed from gas enriched to metallicities $Z>10^{-5} \, Z_\odot$.  Stars above this threshold are drawn from a \citet{Kroupa2001} IMF with $M_{\rm min} = 0.08\,M_\odot$ and $M_{\rm max} = 120 \, M_\odot$.  All Pop II stars with $M \geq 2\, M_\odot$ are tracked as individual particles, while all stars $M<2 \, M_\odot$ formed in a single star formation event are aggregated into a single particle to reduce computational costs. This approximation is justified by their lack of significant feedback on the timescale of the simulation.  We adopt stellar radii, effective temperatures, surface gravity, lifetimes, and length of the asymptotic giant branch (AGB) phase from the PARSEC stellar evolution code \citep{Bressan2012, Tang2014}.

\subsection{Stellar Feedback}

We follow multi-channel stellar feedback from each of our stars including CCSNe (both Pop III and Pop II), SNIa, AGB and massive stellar winds, and ionizing radiation.  Mass and energy feedback from both SNe and massive star winds are deposited over a 2~pc radius sphere centered on the particle.  For both SNe and winds, we only include the thermal energy deposition \citep[see][for details]{AeosMethods}.

CCSNe explode with an energy of $10^{51}$ erg and occur at the end of the lifetime of Pop III stars with $10 \, M_\odot < M_* < 100 \, M_\odot$ and Pop II stars with $8 \, M_\odot < M_* < 25 \, M_\odot$.  We assume that Pop II stars with $M_*>25\, M_\odot$ collapse directly to a black hole with no feedback, as do all stars above the Pop III CCSNe range. Although the Aeos20 simulation has stars in the range of 140--260~$M_\odot$, which are expected to undergo pair instability SNe, we do not include them in this version of \Aeos due to their rarity and the speculative nature of their existence.  Pop II stars with $3 \, M_\odot < M_* < 8 \, M_\odot$ undergo SNIa.  We adopt a delay time distribution from \citet{Ruiter2011} that sums four different channels of SNIa.

\subsection{Stellar Yields}
We track detailed chemical yields from both Pop III and Pop II stars, following H and He, along with 10 metals: C, N, O, Na, Mg, Ca, Mn, Fe, Sr, and Ba, each of which acts as one of the primary tracers for one or more of our nucleosynthetic channels. The $\alpha$ elements O, Mg, and Ca are primary tracers of Pop II CCSNe and short timescale processes, while iron-peak elements Mn and Fe are dominantly produced by SNIa and trace longer timescales; their relative abundances to the $\alpha$-elements act as tracers of star formation history.  Meanwhile, N, Na, Sr, and Ba trace enrichment from AGB stars.  In particular, Sr is a first peak \textit{s}-process element, tracing low-mass ($<4 \, M_\odot$) AGB stars, while Ba is a second peak \textit{s}-process element, tracing higher-mass (4--8~$M_\odot$) AGB stars due to higher neutron densities. [C/Fe] is an important tracer of Pop III CCSNe enrichment.

Stellar yields for Pop III CCSNe with 10--100~$M_\odot$ are adopted from \citet{HegerWoosley2010} using their standard mixing parameter of 0.1.  For CCSNe and winds of massive Pop II stars, we adopt yields from \citet{Limongi2018}, and use a population-averaged mixture model of the different stellar rotations from \citet{Prantzos2018}.  AGB wind yields are adopted from \citet{Cristallo2015}.  Finally, we adopt a single abundance pattern for all SNIa from \citet{Thielemann1986} and opt to instead track the fractional contribution from different channels of SNIa.  This allows us to arbitrarily rescale the abundances from different channels in post-processing.

\section{Comparing Stellar and Gas Abundances} \label{sec:MD}
In order to quantify how well Pop III yields are mixed into star-forming gas and over what scales stellar abundances could be considered homogeneous, we need a metric that not only describes the differences between multidimensional abundance vectors, but also accounts for correlations among elements to minimize the dependence of the difference on the chosen set of elements.  To this end, we use the Mahalanobis distance:
\begin{equation} \label{eq:MD}
    d_{\rm M} = \sqrt{(\textbf{a}_{\rm star} - \textbf{a}_{\rm gas})^{\rm T} \, \textbf{C}^{-1} \, (\textbf{a}_{\rm star} - \textbf{a}_{\rm gas})}
\end{equation}
where $\textbf{a}_{\rm star}$ is a vector of abundances [X/Fe] for a single star, $\textbf{a}_{\rm gas}$ is a vector of abundances [X/Fe] for some amount of gas, and \textbf{C} is the covariance matrix between abundances [X/Fe] derived from all SNe and wind yields injected into the box by the end of the simulation.\footnote{We opt to use a single covariance matrix rather than one that changes over time as the changing variance over time for individual abundances may strongly artificially change $d_{\rm M}$ in a case where the stellar and gas abundance vectors do not otherwise change.  However, we find that using a fixed versus variable covariance matrix does not change the qualitative results nor does it significantly change the quantitative results.  We have also explored covariance matrices derived directly from our yield table (which overweights massive stars), and halo-by-halo, both of which produce very similar results to using yields across the full box.}  Effectively, $d_{\rm M}$ generalizes the Euclidean distance by normalizing for scale and accounting for variable correlations via the inverse covariance matrix.

Notably, we chose to compare abundance vectors [X/Fe] rather than [X/H] or total element mass.  While it could be argued that [X/H] would be the best independent measure of how each element differs between the gas and stellar abundances, [X/H] is dependent on the dilution factor which, when determining the covariance matrix using yields, is impossible to determine without knowing the H content of the volume the yields mix into.  Total element mass is similarly dependent on how diluted it becomes, and we would not expect the mass of any element in gas to remain the same after it becomes locked up in stars during star formation.  We argue that in \Aeos, we do not have to account for inhomogeneous dilution that varies by element because each component mixes into the gas in more or less the same way due to the dominant contribution from SNe at this time \citep{Mead2025a}.  [X/Fe] remains the best choice to compare abundance vectors as each component of the mass fraction is dependent on the yield from some stellar feedback event.

To provide a sense of scaling for $d_{\rm M}$, we examine three simplistic cases: if all abundances differ by 0.01 dex, 0.1 dex, and 0.2 dex.  These numbers represent the average abundance uncertainties for small abundance surveys employing differential abundance analysis \citep[e.g.][]{Bedell2018}, the average abundances uncertainties for large-scale surveys \citep[e.g.][]{apogee_overview,GALAH,LAMOST}, and the upper limits on abundance uncertainties for large-scale surveys, respectively.  Using the \Aeos covariance matrix, this results in $d_{\rm M} = 0.076, \, 0.76, \,$ and $ 1.5$, respectively.

\section{Results} \label{sec:results}
\subsection{Inhomogeneous Mixing in Halos}

We study the extent of metal mixing in \Aeos halos by comparing the stellar abundances of Pop II stars in the simulation for seven abundances [X/Fe] where X is: C, N, O, Na, Mg, Ca, and Mn\footnote{Although \Aeos also includes Sr and Ba, these are not included here as they have not formed in significant amounts in the simulation we are analyzing in this paper.} against three simplified models for the gas abundances: 

\begin{enumerate}
    \item Closed Box Assumption: The abundance [X/Fe] within $r_{\rm vir}$\footnote{The abundance [X/Fe] within the virial radius is calculated by taking the relative number density of element X to Fe within the virial radius, relative to solar values.}, calculated assuming the retention of all yields from SNe and winds within $r_{\rm vir}$ and assuming instantaneous homogeneous mixing of said yields;
    \item Open Box Assumption: The abundance [X/Fe] within $r_{\rm vir}$, calculated assuming homogeneous mixing of all gas phase metals present within $r_{\rm vir}$ and not assuming retention of yields;
    \item Local Cell Assumption: The abundance [X/Fe] within the single gas cell in which the star resides.
\end{enumerate}

\begin{figure}
    \centering
    \includegraphics[width=\linewidth]{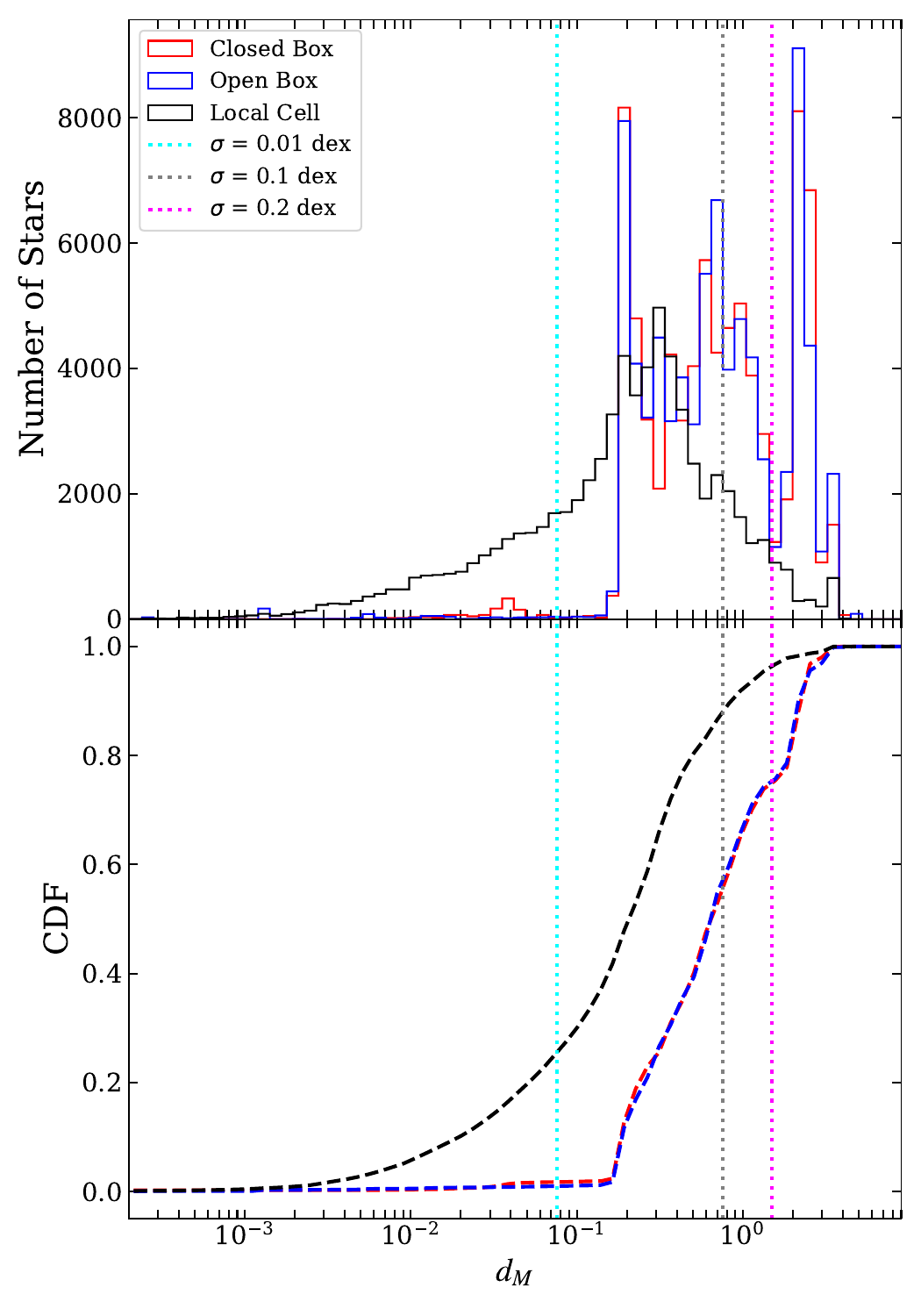}
    \caption{\textit{Top}: Distribution of $d_{\rm M}$, the multidimensional distance between stellar and gas abundance vectors modulated by covariances among element abundances of yields from all chemical feedback, for Closed Box {\em (red)}, Open Box {\em (blue)} and Local Cell {\em (black)} models for gas abundances (see text).  The Closed and Open models have similar distributions in $d_{\rm M}$, whereas the Local Cell model peaks at lower $d_{\rm M}$ and has a more elongated tail, indicating a closer alignment between stellar and gas abundances for this model.  Vertical dotted lines represent the $d_{\rm M}$ if we assume typical observational uncertainties (see Section \ref{sec:MD})---that is, if all abundance differences in our calculation of $d_{\rm M}$ were 0.01 {\em (cyan)}, 0.1 {\em (gray)}, or 0.2 dex {\em (magenta)}, to give a sense of scale for $d_{\rm M}$.  \textit{Bottom}: Cumulative distribution function of $d_{\rm M}$ using the different gas models.}
    \label{fig:MD_closed-leaky-local}
\end{figure}

\begin{figure*}
    \centering
    \includegraphics[trim={1cm 0 1cm 0},clip,width=\linewidth]{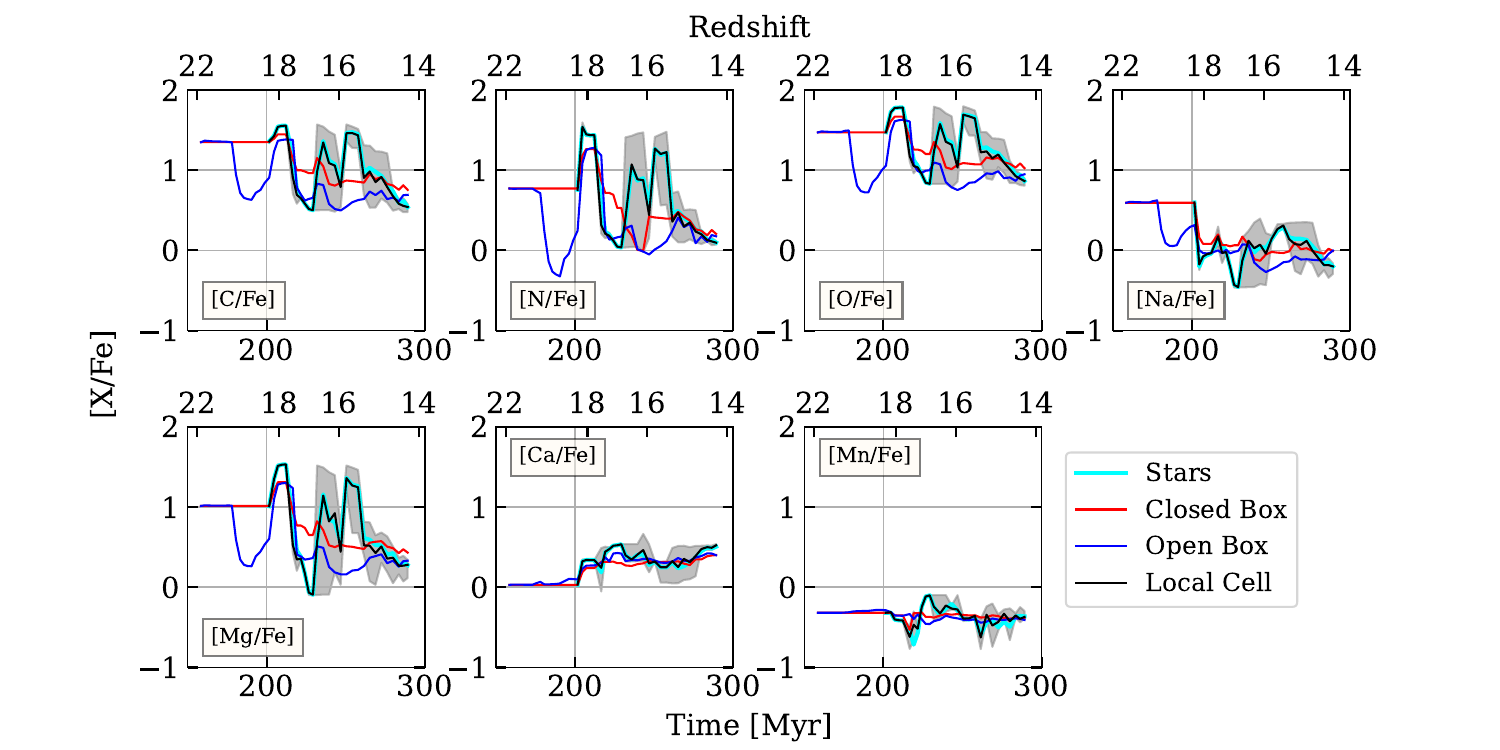}
    \caption{Evolution following the start of Pop III stellar feedback of the average abundance [X/Fe] for all Pop II stars \textit{(cyan)} and the assumptions of Closed Box \textit{(red)}, Open Box \textit{(blue)}, and Local Cell \textit{ (black)} gas abundances, with the 16th to 84th percentile spread of Local Cell abundances \textit{(gray)}.  The stellar and Local Cell average abundances overlap across redshift, whereas the Closed and Open Box average abundances deviate significantly from stellar. Of the abundances we use to calculate $d_{\rm M}$, those of C, N, O, and Mg appear to deviate most significantly for the Closed and Open Box, suggesting these elements contribute most to higher $d_{\rm M}$ values.
    }
    \label{fig:abund_sum}
\end{figure*}

Under each assumption, stellar abundances are compared to the corresponding gas abundance at the time of the star’s first appearance in a simulation snapshot.  Due to the finite snapshot cadence of the simulation, note that a given star may have formed any time between the previous snapshot and the one in which it appears, which allows the gas to evolve in that time frame. This means that, in many cases, stellar abundances are being compared to gas abundances that have evolved away from the state they were in at the time of formation.  The typical snapshot cadence is $\sim$2--4~Myr (on average, 3 Myr).  However, because, for an individual star, the gas abundance will have the same amount of time to evolve, regardless of which assumption we use for the abundance, and the probability that a star will form at any point between snapshots is uniform, we determine that qualitative comparisons between distributions of $d_{\rm M}$ are still valid.

Figure~\ref{fig:MD_closed-leaky-local} compares the distribution of $d_{\rm M}$ of each stellar abundance relative to each of the above gas abundance vectors in the snapshot the star first appears in.  The top panel shows the histogram (i.e. proportional to differential distribution) of $d_{\rm M}$, whereas the bottom panel shows the cumulative distribution.  We will use the cumulative distribution in future versions of this figure, but include the histogram to illustrate how the distributions compare, namely to emphasize that shallower slopes at high $d_{\rm M}$ correspond to a decrease in the number of high $d_{\rm M}$ values for a particular gas abundance comparison.

From Figure \ref{fig:MD_closed-leaky-local}, it is immediately clear that stellar and gas abundance vectors are more closely aligned when adopting the Local Cell model in the snapshot immediately following the star's formation, than the gas abundance of the galaxy when using either the Closed or Open Box assumptions.  This is apparent from both the elongated tail at lower $d_{\rm M}$ and the lower peak of $d_{\rm M}$ for the Local Cell distribution.  The higher distribution of $d_{\rm M}$ for the Closed and Open box assumptions underscores that instantaneous and homogeneous mixing of yields across the virial radius of a galaxy is a poor approximation for describing how chemical enrichment proceeds.

Furthermore, the Closed and Open Box assumptions have similar final distributions, suggesting that the gas abundance vectors in halos are similar no matter which of these assumptions is invoked.  This is expected, as CCSNe are the dominant nucleosynthetic source at this time, so all individual elements behave and distribute similarly as they are mixed into the ISM and intergalactic medium \citep{Mead2025a}.

We can gain a sense of which elements impact $d_{\rm M}$ the most by plotting the average abundance of each element for each abundance measure used, as shown in Figure \ref{fig:abund_sum}.  From the figure, it is clear that the stellar abundances are matched best when assuming the averaged Local Cell abundances. The Closed and Open Box assumptions deviate significantly.  We also see common evolution between elements: C, N, O, Na, and Mg demonstrate similar qualitative trends, which we attribute to their common origins in CCSNe at this time, meaning that they mix into the surrounding gas in a similar fashion.  Furthermore, C, N, O, Na, and Mg all show a sharp drop in abundance for the Open Box assumption around 180 Myr, which can be attributed to metals being expelled beyond the virial radius \citep[see][for additional exploration of both of these effects]{Mead2025a}.  Each of these elements demonstrates larger deviations between the Closed/Open Box assumptions and the Local Cell model, indicating that they contribute most to higher $d_{\rm M}$ over [Ca/Fe] and [Mn/Fe].

We again note that, although Sr and Ba are tracked in \Aeos, we do not include [Sr/Fe] or [Ba/Fe] in our calculations of $d_{\rm M}$.  For Sr and Ba, the average stellar abundances vary by up to 6 orders of magnitude on short timescales, which is not seen in even the Local Cell abundances.  This occurs because, at this stage, Sr and Ba are primarily produced in CCSNe in trace amounts; any increase in the amount of Sr and Ba induces an order of magnitude changes in their abundances relative to Fe.  We expect that once AGB stars become the dominant producers of Sr and Ba, these variations will level out.

The Closed/Open Box and the Local Cell assumptions effectively represent two extremes for the gas abundances: the abundance of metals across the whole halo versus the immediate, local gas from which the star formed, respectively.  However, somewhere in between these two extremes must be a metal mixing scale out to which metals are well-mixed while the gas abundances still remain close to the stellar abundances (in $d_{\rm M}$).  To investigate where this minimum lies, we calculate the gas abundance within concentric spheres of 0.01, 0.025, 0.05, 0.075, 0.1, 0.5, 1, and 5 physical kpc\footnote{The halo virial radius $r_{\rm vir}$ of our Pop II galaxies ranges from 0.1--0.8~kpc, with an average of 0.3~kpc.} around each star-hosting cell at each snapshot, similarly to what was done for the Closed and Open Box assumptions.  We then track how the abundance of each element, relative to Fe, evolves both spatially and temporally.

\begin{figure}
    \centering
    \includegraphics[trim={0cm 0cm 0cm 0cm},clip,width=\linewidth]{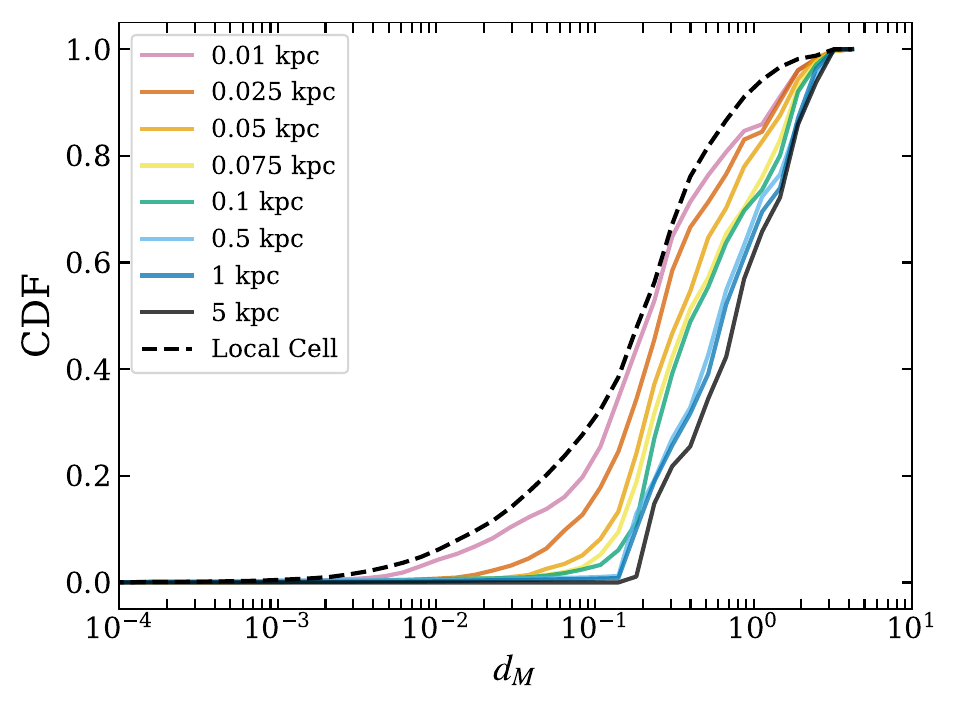}
    \caption{Cumulative distribution of $d_{\rm M}$ using gas abundances calculated assuming instantaneous and homogeneous mixing across spheres of some radius $R$ centered on sites of star formation, along with the Local Cell model for comparison.  Approximating instantaneous homogeneous mixing across spheres up to $R=0.1$ kpc results in a tail of small $d_{\rm M}$, with smaller volumes reaching smaller $d_{\rm M}$, as well as smaller volumes having fewer stars at high $d_{\rm M}$.}
    \label{fig:MD_spheres_hist}
\end{figure}

Figure \ref{fig:MD_spheres_hist} shows the cumulative distribution of $d_{\rm M}$ for each concentric sphere around each star in the snapshot following the time at which it forms.  There is an apparent gradient among the lowest $d_{\rm M}$ values, where the gas abundances of smaller volumes are more closely aligned with the stellar abundances.  The thick tail of low $d_{\rm M}$ exists for spheres up to $0.1$ kpc, implying that within this scale, gas may be mixed homogeneously. In the differential distribution histogram these appear as skewed Gaussians in log-space, similar to the Local Cell assumption in Figure~\ref{fig:MD_closed-leaky-local}, with a peak at 2--3$\times10^{-1}$. Larger spheres show an approximately log-uniform distribution at $10^{-1} < d_{\rm M}< 5$, similar to the Closed and Open Box assumptions in Figure \ref{fig:MD_closed-leaky-local}.

\begin{figure}
    \centering
    \includegraphics[trim={0.15cm 0.5cm 0cm 0.2cm},clip,width=\linewidth]{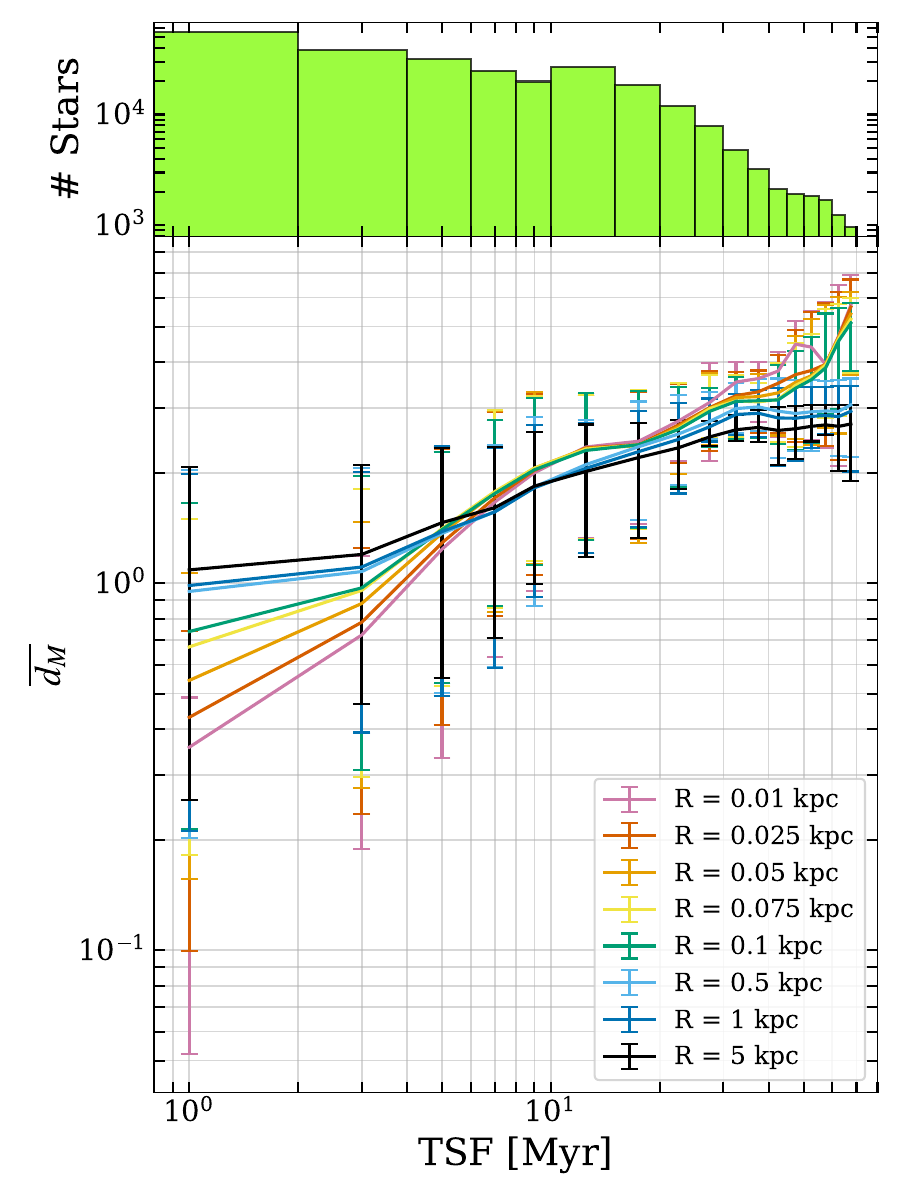}
    \caption{\textit{Top}: Number of stars per time bin.  \textit{Bottom}: $\overline{d_{\rm M}}$ along with 16th to 84th percentile spread of $d_{\rm M}$ over TSF assuming instantaneous and homogeneous mixing across spheres of radius $R$. (The zero-point of TSF is the time at which each star forms.)  Prior to $\sim 7$~Myr after a star forms, smaller spheres have smaller $\overline{d_{\rm M}}$.  The $\overline{d_{\rm M}}$ for each sphere intersects at $\sim7$ Myr, suggesting $\tau_{\rm chem} = 7$ Myr.  After 7 Myr, the $\overline{d_{\rm M}}$ of spheres $R\leq0.1$ kpc continues to increase dramatically, consistent with rapidly changing local gas, whereas the $\overline{d_{\rm M}}$ of spheres $R\geq0.5$ kpc flattens, consistent with expectations for an IMF-averaged gas abundance.}
    \label{fig:MD_spheres_TSF}
\end{figure}

It is clear from Figures \ref{fig:MD_closed-leaky-local} and \ref{fig:MD_spheres_hist} that the gas abundances are inhomogeneous across the extent of the halos.  However, this is only at the time of star formation, so it does not capture how the gas evolves following star formation.  Therefore, we also examine how $d_{\rm M}$ evolves within each sphere around a star following its formation.  Figure \ref{fig:MD_spheres_TSF} shows $\overline{d_{\rm M}}$, the value of $d_{\rm M}$ averaged over the time since star formation (TSF) within each sphere (the zero-point of the TSF is the time at which the star forms). For any given TSF bin, stars are drawn from different cosmic times, and not a uniform time, redshift, or snapshot.

Clear trends among the averages immediately arise from this.  First, near the time of star formation, we see that $\overline{d_{\rm M}}$ decreases with decreasing sphere radius (as also implied by Figure \ref{fig:MD_spheres_hist}).  While $\overline{d_{\rm M}}$ of the 0.5, 1, and 5 kpc spheres are comparatively flat, showing only a modest increase (likely due to the spheres being large enough to encompass enough gas to be representative of IMF-averaged yields), smaller spheres demonstrate an increase in $\overline{d_{\rm M}}$. The $\overline{d_{\rm M}}$ for each sphere intersects at $\sim 7$ Myr, suggesting an average timescale for significant chemical evolution $\tau_{\rm chem}\sim 7$ Myr, where the gas in spheres smaller than $R = 0.1$ kpc has altered sufficiently to be as distant from the original local abundance as IMF-averaged gas, either due to yields injected by nearby stars, or as the star drifts beyond the gas from which it formed.  Moreover, the values of $\overline{d_{\rm M}}$ of spheres $R \leq 0.1$ kpc continue to rise together after this mixing timescale. This rise demonstrate that the gas within $R \leq 0.1$ kpc changes more significantly and more rapidly than IMF-averaged gas $R \geq 0.5$ kpc. This further demonstrates that there is a spatial mixing scale $\sim 0.1$ kpc over which the gas is similarly mixed.

\begin{figure}
    \centering
    \includegraphics[trim={0.5cm 2cm 1cm 3.5cm},clip,width=\linewidth]{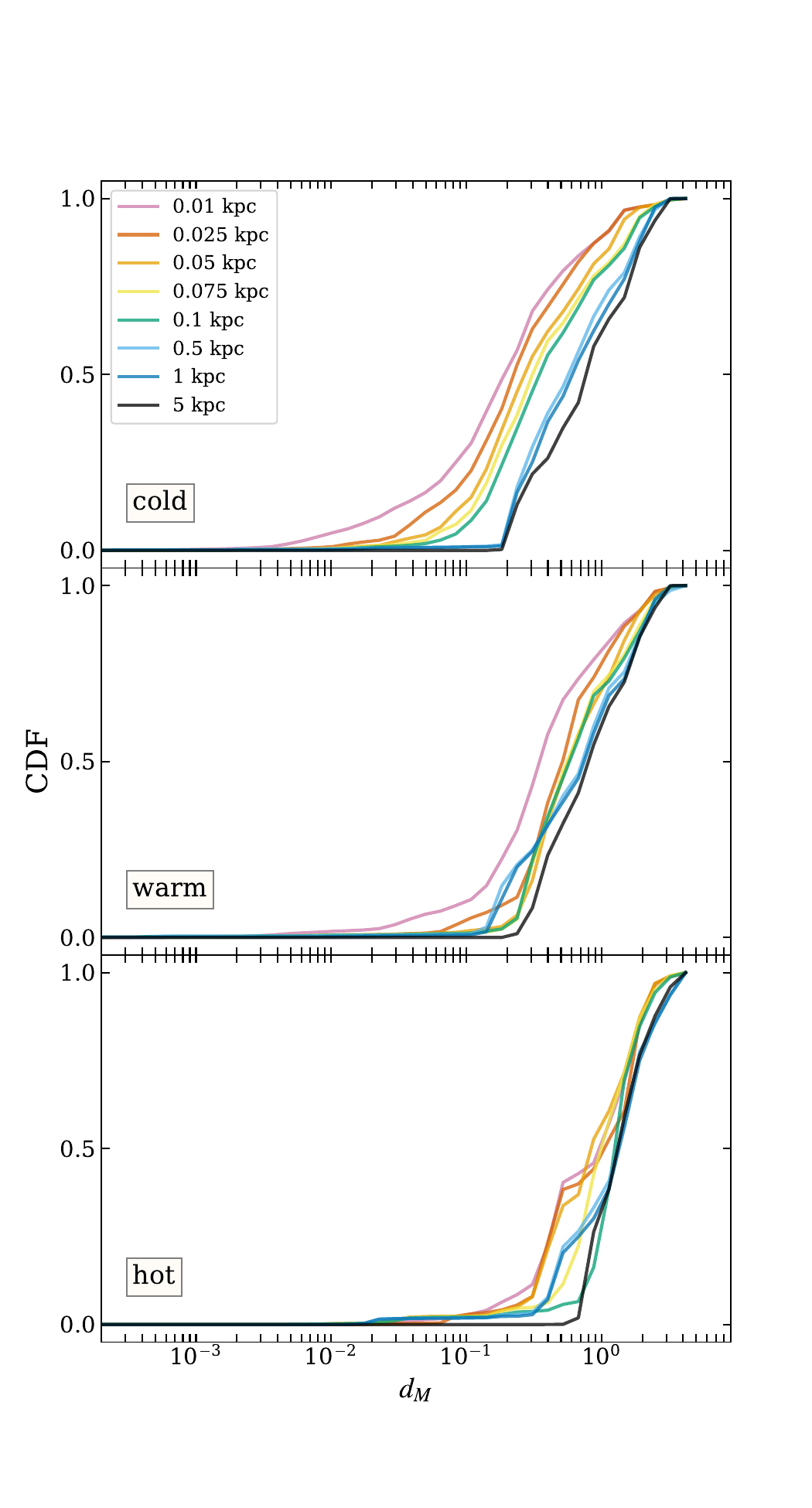}
    \caption{Cumulative distribution of $d_{\rm M}$ assuming instantaneous and homogeneous mixing of gas across spheres with radius $R$ given in the legend, separated by gas temperature (see text).  As gas gets progressively warmer, the tail of small $d_{\rm M}$ shrinks, indicating that cold gas in spheres with $R\leq0.1$ kpc has gas abundances most similar to stellar abundances.  Warm gas shows a small tail for spheres $R\leq0.025$ kpc, whereas hot gas distributions overlap for every size sphere.  This clearly demonstrates that cold gas is most representative of young stellar abundances while hot gas is never representative of them.}
     \label{fig:gas_phase_hist}
\end{figure}

\begin{figure*}
    \centering
    \includegraphics[trim={2.5cm 1.5cm 3cm 3.9cm},clip,width=\linewidth]{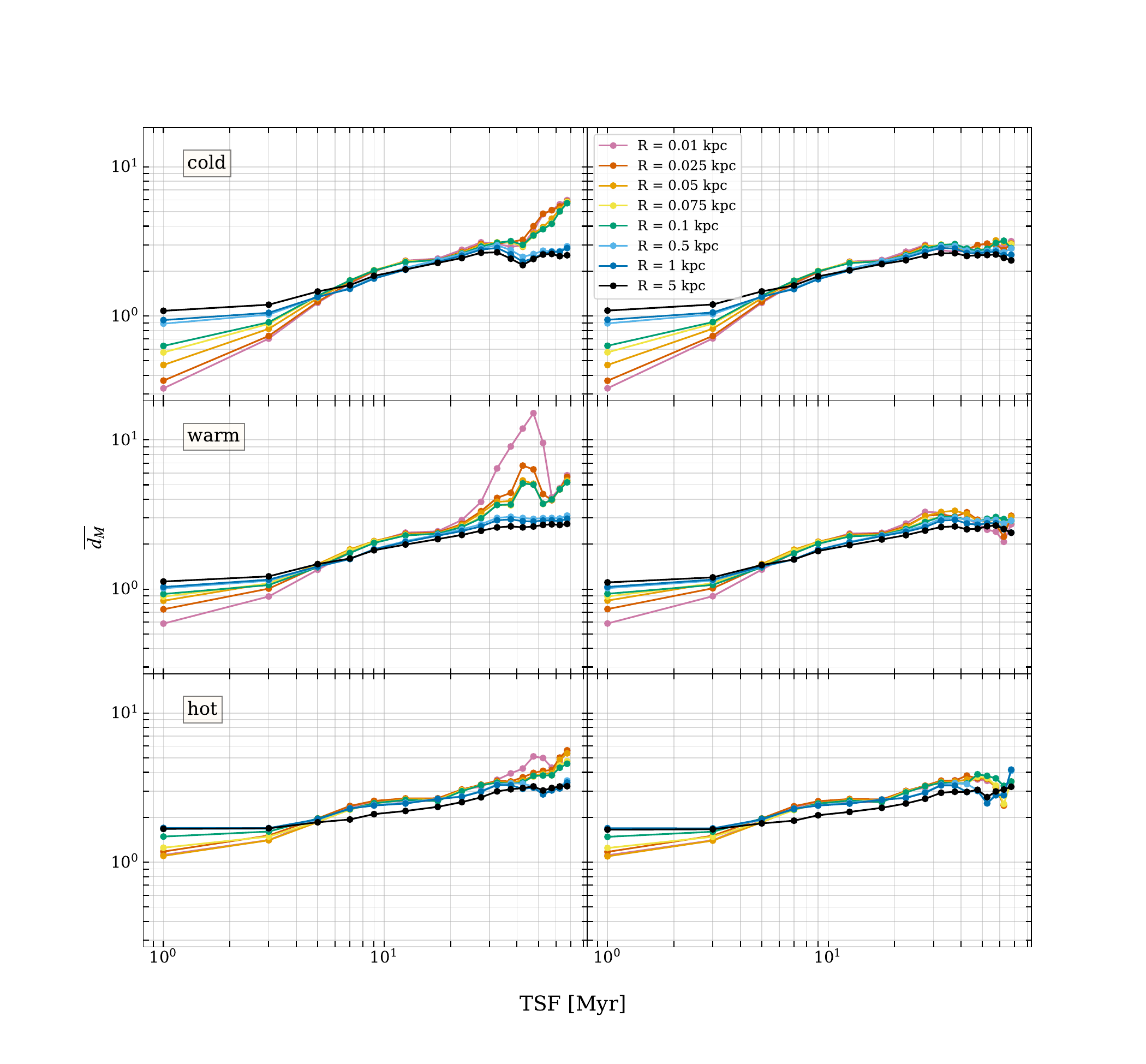}
    \caption{\textit{Left}: $\overline{d_{\rm M}}$ over TSF by gas temperature for all halos in the sample.  One halo, Halo 159508, dominates early Pop II star formation, having an outsized impact on $\overline{d_{\rm M}}$ at high TSF.  Unique circumstances in Halo 159508 contribute to an anomalous bump in $\overline{d_{\rm M}}$ in the warm gas at $30 < {\rm TSF} < 60$ Myr, as well as the increase in $\overline{d_{\rm M}}$ at large TSF in the cold and hot gas. \textit{Right}: Same as left column but with Halo 159508 removed.  General trends remain, but without the severe increase in $\overline{d_{\rm M}}$ at large TSF in the cold gas.  Although less pronounced, there is still a reversal in sphere size for largest $\overline{d_{\rm M}}$ after 10 Myr, where spheres $R\leq0.1$ kpc have higher $\overline{d_{\rm M}}$.}
    \label{fig:gas_phase_TSF}
\end{figure*}

\subsection{Mixing in Different Gas Phases}

Metals will mix differently in different thermodynamic phases of the gas with different sound speeds.
In particular, it is interesting to distinguish how representative cold gas is of stellar abundances compared to hot gas that may be associated with SN ejecta.  We repeat the analysis done earlier, using the same concentric spheres around sites of star formation, but divide the gas into cold $(T<1{,}000 \, \rm K)$, warm $(1{,}000<T<20{,}000 \, \rm K)$, and hot $(T>20{,}000 \, \rm K)$ gas.

Similarly to Figure \ref{fig:MD_spheres_hist}, Figure \ref{fig:gas_phase_hist} shows the cumulative distribution of $d_{\rm M}$, but now by gas phase.  While most values of $d_{\rm M}$, regardless of sphere size, remain larger than $10^{-1}$, there is a noticeable shift in the distribution at lower $d_{\rm M}$ in each phase of gas.  Namely, cold gas demonstrates a pattern most similar to that seen in Figure \ref{fig:MD_spheres_hist}, where spheres of $R \leq 0.1$ kpc have a thick tail of low $d_{\rm M}$, with smaller spheres reaching lower $d_{\rm M}$.  This tail begins to disappear in warm gas, with only a small tail for spheres of $R \leq 0.025$ kpc, before it disappears completely in hot gas, leaving the distribution of $d_{\rm M}$ among the different size spheres almost indistinguishable. This is expected, since stars form from cold gas (although without a density criterion the cold gas filter may also select non–star-forming gas) so it is sensible that some cold gas around sites of star formation should have abundances very similar to newly formed stars.

Conversely, the higher $d_{\rm M}$ values for the warm and hot gas abundances mean there is essentially never a time at which the warm or hot gas is representative of individual stellar abundances.  Uniformly large $d_{\rm M}$ values could be expected for two reasons: if the hot gas mixes efficiently \citep{Avillez2002,Emerick2018b,Emerick2020a} and thus rapidly diverges from stellar abundances; or if SNe recently enriched the local hot gas, potentially from the very star being compared to, considering the snapshot cadence. Together, these results suggest that ejecta from SNe do not mix instantaneously or homogeneously between phases of gas.

We continue our analysis of $d_{\rm M}$ in different gas phases by comparing $\overline{d_{\rm M}}$ in different spheres over TSF, as in Figure \ref{fig:MD_spheres_TSF}.  The left column of Figure \ref{fig:gas_phase_TSF} shows $\overline{d_{\rm M}}$ for each gas phase in each sphere.  Within $\sim7$ Myr, each progressively larger sphere shows a rise in $\overline{d_{\rm M}}$ that also rises with gas temperature, and for spheres of $R > 0.1$ kpc, $\overline{d_{\rm M}}$ flattens at high TSF.  This behavior again suggests that metals are not mixed homogeneously across gas phases, with cold gas within 7 Myr of star formation being the most similar in abundances to stars.  Within each sphere, the cold and warm gas show similar evolution in $\overline{d_{\rm M}}$ with TSF, and the hot gas sits at a consistently higher $d_{\rm M}$ out to about 20 Myr for all spheres, and after 20 Myr for spheres $R\geq0.5$ kpc.  The similarities between the cold and warm gas become stronger after 10 Myr, and in spheres of $R > 0.1$ kpc, with the exception of a large bump in $\overline{d_{\rm M}}$ of warm gas for $30 < {\rm TSF} < 60$ Myr that is due to a single halo, Halo 159508, that dominates most of the star formation at early times, thus producing most of the star particles with large TSF  in snapshots at late times.

We remove Halo 159508 from our analysis in the right column of Figure \ref{fig:gas_phase_TSF}.  Removing this halo not only removes the bump in $\overline{d_{\rm M}}$ in warm gas, but also removes the steep increase in $\overline{d_{\rm M}}$ at late TSF in spheres $R\leq0.1$ kpc for each phase of gas.  However, the trends between sphere sizes remain, namely that smaller spheres at large TSF generally retain a higher $\overline{d_{\rm M}}$.  How rapidly local gas abundances change compared to stellar abundances of stars formed more than $\sim 40$ Myr prior is clearly environment dependent, and likely is a result of stellar motion into regions of gas far from their formation sites, ongoing star formation, and SN feedback.  Though the impact of removing Halo 159508 is pronounced due to its dominant contribution to the oldest stars, this emphasizes that each halo has a unique shape in $\overline{d_{\rm M}}$ versus TSF. However, halos generally exhibit the same qualitative trends, with smaller volumes having lower $\overline{d_{\rm M}}$ until about 5--10~Myr after star formation, at which point this reverses with smaller volumes having higher $\overline{d_{\rm M}}$.

\section{Discussion and Conclusions} \label{sec:conc}
In this work, we use the \Aeos simulations to test the assumption of instantaneous and homogeneous mixing of yields across different gas phases in early ($z>14$) minihalos with masses $10^5 \lesssim M_h \lesssim 10^7 \, M_\odot$ and to identify the temporal and spatial scales of mixing.  We use the Mahalanobis distance---similar to the Euclidean distance between vectors but adjusted for covariances between variables---to quantify the differences between stellar and gas abundances.  As stellar abundances presumably record the state of the gas in a particular location at a particular instant, we study how the gas enriched by Pop III SNe evolves spatially and temporally relative to each instance of Pop II star formation.  We quantify the gas abundance in various ways to test the assumption of instantaneous and homogeneous mixing.  We test three models: (i) a Closed Box model with full metal retention in a halo; (ii) an Open Box model with only the metals actually retained, which both assume instantaneous and homogeneous mixing across a halo out to the virial radius; and (iii) a Local Cell model, which only accounts for the gas in the cell in which the star formation occurred. We also repeat the analysis for a series of concentric spheres of various radii, which tests instantaneous and homogeneous mixing across increasingly larger spheres from 0.01 kpc, well within the virial radius, out to 5 kpc, well beyond the virial radius of any Pop II star forming halo in our simulations.  From our analysis, we conclude the following:
\begin{enumerate}
    \item Metals do not mix instantaneously or homogeneously across halos, whether or not we assume complete metal retention by halos. We measure gas abundances immediately around star forming regions and find significant differences with stellar abundances compared to the abundances that are expected when assuming that enriched material instantaneously and homogeneously mixes within the virial radius of the halo.
    \item Metals are mixed to a point where abundances approximated with instantaneous and homogeneous mixing within different spherical volumes are equidistant in Mahalanobis distance $d_{\rm M}$ within $\sim 7$ Myr.  This implies a mixing timescale for significant chemical evolution to occur $\tau_{\rm chem} \sim 7$ Myr.
    \item Metals are mixed homogeneously within $\sim 0.1$ kpc.  This effect is particularly apparent at large times after star formation because gas in small volumes surrounding sites of star formation changes more rapidly than gas abundances averaged over large spheres.
    \item Abundances in cold gas are most similar to stars within $0.1$ kpc around sites of star formation, reaching lower $d_{\rm M}$.  Conversely, hot gas demonstrates comparatively high $d_{\rm M}$ regardless of the volume over which instantaneous and homogeneous mixing is assumed, indicating a consistent and significant deviation from stellar abundances that is likely due to some combination of efficient mixing and local enrichment.  Finally, warm gas is moderately representative of stellar abundances within $0.025$ kpc around a site of star formation.
\end{enumerate}

Metal mixing scales have implications for both theoretical and observational work.  The instantaneous and homogeneous mixing assumption has long persisted in analytical work due to its simplicity and tractability.  While this has been sufficient to describe most stellar abundances observed at [Fe/H] $\geq -2$, outliers that suggest inhomogeneous mixing persist, and data from the most metal-poor stars strongly supports inhomogeneous mixing \citep[][]{Ryan1996,McWilliam1997,Francois2003}.  In order to match these observations and explain the scatter that is observed in abundances with age and overall metallicity [Fe/H] \citep[e.g.][]{Hill2019, Ji2020, Mead2024}, analytical models must adopt new ways to account for spatial and temporal mixing scales \citep[e.g.][]{Krumholz2018,Krumholz2025}.

Spatial and temporal mixing scales also impact chemical tagging, a central pursuit of galactic archaeology.  Chemical tagging relies on there being enough independent chemical dimensions to uniquely identify conatal stars under the assumption that stars that are born together from the same gas cloud should have the same abundances.  Our work here suggests that this is a valid assumption: we find a spatial mixing scale of about $0.1$ kpc, and the average size of a giant molecular cloud ranges from 20--100 pc \citep[][]{Solomon1979,Miville-Deschnes2017}. Although we see a mixing timescale of $\sim7$ Myr, this is the timescale over which gas interior to $~0.1$ kpc mixes sufficiently across the volume to be as distinct from the stellar abundances as the IMF-averaged gas abundances. This does not imply that stars forming within 7 Myr of each other have indistinguishable abundances.

Under the assumption that stellar abundances are records of the gas abundances from which they formed, we can ask the question: given observational uncertainties, what is the minimum $d_{\rm M}$ between two observed stars that would indicate they formed from different gas clouds?  We can do a simple calculation of $d_{\rm M}$ for the elements in \Aeos using a typical survey observational uncertainty of 0.1 dex so $\textbf{a}_{\rm star1} - \textbf{a}_{\rm star2} = [0.1, \, \, \, 0.1, \, \, \, 0.1, \, \, ...]$.  Using the covariance matrix from \Aeos, we find a minimum $d_{\rm M} = 0.76$.  Notably, this value is higher than a majority of $d_{\rm M}$ at the time of star formation in volumes $R \leq 0.1$ kpc (e.g. Figures \ref{fig:MD_closed-leaky-local}, \ref{fig:MD_spheres_hist}, \ref{fig:gas_phase_hist}), but smaller than around half of the values for volumes $R\geq 0.5$ kpc.  With these uncertainties, a value of $d_{\rm M} < 0.76$ could constrain the formation time of two stars to be within $\sim 10$ Myr of each other (below the 16th percentile spread in Figure \ref{fig:MD_spheres_TSF}).

Using an upper limit on average abundance uncertainties of 0.2 dex, the minimum $d_{\rm M} = 1.5$, which, we see from Figure \ref{fig:MD_spheres_TSF}, would not allow us to distinguish different gas clouds, but could narrow the range of ages to within 20~Myr of each other.  Conversely, assuming an abundance uncertainty of 0.01 dex, which can be achieved by studies implementing differential abundance analysis \citep[e.g.;][]{Bedell2018}, translates to $d_{\rm MD} = 0.076$, and would nearly uniquely identify stars forming from the same gas cloud with very little contamination from stars forming outside of that gas cloud (Figures \ref{fig:MD_closed-leaky-local} and \ref{fig:MD_spheres_hist}). However, as shown in Figure \ref{fig:MD_spheres_hist}, given than many stars have $d_{\rm M}$ larger than this, such a cutoff would not capture all stars that form from the same gas cloud.  It should be noted, though, that such small uncertainties can at best be achieved from differential abundance analyses tied to the Sun. These have thus far only been applied to metal-rich, solar-type stars and solar twins \citep[e.g.,][]{Bedell2018}. 

In any case, we reiterate that in our current analysis, time steps are spaced 2--4~Myr apart, which is sufficient time for gas abundances to be altered by CCSNe. This would particularly impact stars forming immediately after the previous time step, but 2--4~Myr before the gas abundances are recorded.  \Aeos does have higher time resolution outputs, so a more detailed future analysis could further separate these $d_{\rm M}$ distributions.  An important caveat to these constraints is that they depend on \Aeos having an accurate covariance matrix, as well as on which elements are included in our calculation of $d_{\rm M}$, particularly elements that are independent from the others in our set. Theoretically, the addition of elements that are highly correlated with each other should not greatly impact the value of $d_{\rm M}$ \citep[see arguments from e.g.,][]{Mead2025b,Krumholz2025}.  Nonetheless, our results highlight the inhomogeneity in abundances in the metal-poor regime, and the back-of-the-envelope calculations for $d_{\rm M}$ given expected uncertainties demonstrate a potential method by which simulations and observations can work hand-in-hand to trace the origins of the earliest stars.

\section*{Acknowledgments}
J.M. acknowledges support from the NSF Graduate Research Fellowship Program through grant DGE-2036197. K.B. is supported by an NSF Astronomy and Astrophysics Postdoctoral Fellowship under award AST-2303858. G.L.B. acknowledges support from the NSF (AST-2108470 and AST-2307419, ACCESS), a NASA TCAN award, and the Simons Foundation through the Learning the Universe Collaboration. This research was supported in part by grant NSF PHY-2309135 to the Kavli Institute for Theoretical Physics (KITP).  M.-M.M.L. and E.P.A. acknowledge partial support from NSF grant AST-2307950 and NASA ATP grant 80NSSC24K0935. A.P.J. acknowledges support from NSF grants AST-2206264 and AST-2307599. J.H.W. acknowledges support from NSF grants AST-2108020 and AST-2510197 and NASA grant 80NSSC21K1053. A.F.\ acknowledges support from NSF grant AST-2307436.

The authors acknowledge the Texas Advanced Computing Center at The University of Texas at Austin for providing high-performance computing and storage resources (LRAC-AST20007) that have contributed to the research results reported within this paper.

\bibliographystyle{yahapj}
\bibliography{refs}

\end{document}